\documentclass[
	a4paper, 
	10pt, 
	unnumberedsections, 
	twoside, 
]{LTJournalArticle}
\usepackage{tabularx} 
\usepackage{amsmath}
\usepackage{multirow}%
\usepackage{makecell}

\title{A Graph Neural Network based on a Functional Topology Model: Unveiling the Dynamic Mechanisms of Non-Suicidal Self-Injury in Single-Channel EEG} 

\author{%
	BG Tong\textsuperscript{1}\thanks{Corresponding author: \href{mailto:dackmoon123@126.com}{dackmoon123@126.com}}
}

\date{\footnotesize\textsuperscript{\textbf{1}}Department of Psychiatry, Inner Mongolia People’s Hospital, Hohhot, China}

\renewcommand{\maketitlehookd} {%
\begin{abstract}
\textbf{Objective:} This study aims to propose and preliminarily validate a novel "Functional-Energetic Topology Model" to unveil the underlying neurodynamic mechanisms of Non-Suicidal Self-Injury (NSSI). We explore the utility of Graph Neural Networks (GNNs), an advanced algorithm, to decode interpretable brain functional network patterns associated with NSSI impulses from single-channel Electroencephalography (EEG) signals collected in naturalistic settings.\textbf{Methods:} We recruited three adolescent NSSI patients and collected their EEG data during NSSI-impulsive and non-impulsive periods over approximately one month, using a digital therapeutic paradigm based on a smartphone APP and a portable Fp1 single-channel EEG headband. A GNN model based on our theory, comprising seven functional nodes, was constructed. Its performance was assessed via intra-subject validation (80/20 split) and a rigorous leave-one-subject-out cross-validation (LOSOCV) scheme. Finally, GNNExplainer was employed for explainability analysis to identify key functional pathways.\textbf{Results:} The GNN model exhibited high accuracy in the intra-subject prediction task (average accuracy > 85\%) and achieved performance significantly outperforming random chance in the cross-subject generalization test (average accuracy $\approx$ 73.7\%). The core finding emerged from the explainability analysis: compared to the functional non-NSSI state, the onset of the NSSI state is critically associated with the dysfunction and directional reversal of a key feedback regulatory loop responsible for processing somatic sensations. In the NSSI state, the system not only loses its ability to self-correct based on negative somatic feedback, but its defense mechanism itself appears to enter a state of "ineffective idling".\textbf{Conclusion:} This study successfully demonstrates the feasibility of applying a clinically theory-driven GNN model to sparse, single-channel EEG data for decoding complex mental states. The discovered "feedback loop reversal" model provides a novel, computable, and dynamic perspective for understanding the internal mechanisms of NSSI, opening a promising avenue for the future development of objective biological markers and next-generation Digital Therapeutics (DTx). \\\\
\textbf{Keywords:} Non-Suicidal Self-Injury (NSSI), Graph Neural Network (GNN), Electroencephalography (EEG), Explainable AI (XAI), Computational Psychiatry, Functional Topology
\end{abstract}
}

\begin{document}

\maketitle 


\section{Introduction}

\subsection{The Clinical Landscape and Challenges of Non-Suicidal Self-Injury (NSSI)}

Non-Suicidal Self-Injury (NSSI) is defined as the deliberate, direct destruction or alteration of one's own body tissue without any explicit suicidal intent, with its most common forms including skin cutting, burning, or scratching \cite{reinhardt2025interconnectivity}. Although NSSI is not intended to be fatal, it has become an increasingly severe global public health issue, particularly prevalent among adolescents. Epidemiological surveys indicate that the prevalence of NSSI among adolescents worldwide is above 17\% \cite{he2025impulsivity}, with rates being even higher in some clinical samples. Chronic NSSI not only causes direct physical harm and scarring but is also highly comorbid with various psychiatric disorders such as depression, anxiety, and borderline personality disorder \cite{swannell2014prevalence}, significantly increasing the risk of future suicidal ideation and attempts. This poses a serious and long-term challenge to the mental health and social functioning of adolescents.

However, the core challenge currently faced in clinical practice lies not only in addressing its high prevalence but also in the insufficient understanding of its complex underlying psychological mechanisms. Although existing research widely regards NSSI as a maladaptive emotion regulation strategy used to cope with unbearable negative emotions \cite{han2024impact}, this broad functional explanation fails to fully elucidate a unique and critical phenomenological feature of NSSI: its deep connection with somatization and other Body-Focused Repetitive Behaviors (BFRBs) \cite{gatta2022alexithymia}.

From the perspective of our clinical observations and developmental psychology, we propose that NSSI may not be a simple behavioral problem but rather a dynamic process in a specific neurodevelopmental context where emotion regulation pathways are "short-circuited", resorting to primitive somatic-motor patterns. Specifically, it exhibits features of a "reversed" or "immature" somatization. Unlike classic, implicit somatization primarily governed by the autonomic nervous system (e.g., gastrointestinal distress in adults due to anxiety) \cite{mayer2001depression}, NSSI in adolescents displays an explicit, active, and almost "concretized" characteristic. Adolescents do not passively endure abstract physiological discomfort caused by internal emotions; instead, they "manufacture" a perceptible, concrete physical pain through active bodily actions to anchor, explain, or overwhelm an otherwise ineffable inner turmoil \cite{fabbrini2022self}. As psychoanalytic theory suggests, this behavior can be seen as an immature defense mechanism \cite{xia2023correlation}, using a controllable external harm to replace or account for uncontrollable internal chaos.

Furthermore, NSSI shares significant behavioral homogeneity with BFRBs such as trichotillomania and dermatillomania \cite{mathew2020body}. These behaviors often manifest under emotional distress as difficult-to-control, repetitive self-harming habits, which may collectively point to a specific developmental stage in adolescence characterized by the incomplete differentiation of the emotional and somatosensory-motor systems \cite{xie2025impulse}. In a typical developmental trajectory, individuals gradually learn to separate emotional experiences from physical actions, developing higher-order abstract thinking and linguistic symbols as mediators for emotion regulation \cite{liu2020emotion}. For some adolescents, however, this process of mentalization may be hindered by internal and external factors such as genetics, environment, or psychological trauma \cite{hajek2024efficacy}. They appear to be "stuck" in a "chaotic-integrative" state where emotion and soma remain highly coupled \cite{armey2008comparison}, akin to the early cognitive development described by Piaget, where thought and emotion are heavily dependent on concrete actions \cite{emelianchik2019non}. Consequently, when faced with intense emotional impact, lacking mature internal regulation pathways, they can only regress to this primitive coping style—attempting to process, release, or "complete" the blocked emotional process through an active physical action, i.e., self-injury \cite{hasking2017cognitive}.

In summary, the key to understanding NSSI may lie in unveiling the dynamic transition mechanism from emotional dysregulation to concrete self-injurious behavior. Exploring the unique patterns of brain functional activity during this process is fundamental to providing objective biological markers for the early identification, state monitoring, and effective intervention of NSSI. Therefore, the starting point of this research is to attempt to construct a computational model capable of capturing and explaining this dynamic process.

\subsection{Existing Research and Limitations in the Neurobiological Mechanisms of NSSI}

In recent years, with the advancement of neuroimaging and electrophysiological techniques, significant progress has been made in exploring the neurobiological mechanisms of NSSI, identifying a series of core brain regions and networks related to emotion regulation \cite{wang2025neural}. Functional Magnetic Resonance Imaging (fMRI) studies have consistently found that individuals with NSSI exhibit hyperactivation of the limbic system, centered around the amygdala, when faced with negative emotional stimuli, reflecting their heightened emotional reactivity \cite{branas2021neuroimaging}. Concurrently, brain regions responsible for higher-order cognitive control and emotional inhibition, such as the Prefrontal Cortex (PFC)—particularly the dorsolateral prefrontal cortex (dlPFC) and the ventromedial prefrontal cortex (vmPFC)—show relatively diminished activity and functional connectivity with the limbic system. This imbalanced pattern, characterized by enhanced "bottom-up" emotional drive and insufficient "top-down" cognitive control, is considered a core feature of NSSI neuro-functionality. Furthermore, abnormal activation of the insula, a key node connecting emotional experience with internal bodily sensations (interoception), suggests that the emotional pain of NSSI patients is more readily experienced as intense and unbearable somatic discomfort \cite{zhou2025study}. At the electrophysiological level, Electroencephalography (EEG) studies, using metrics like Event-Related Potentials (ERPs), have also revealed abnormalities in the processing of emotional information in individuals with NSSI, such as a stronger Late Positive Potential (LPP) in response to negative stimuli, further corroborating the excessive consumption and difficulty in regulating their emotional resources \cite{zhao2025impaired}.

Although the aforementioned studies have provided us with a "neural snapshot" of NSSI, identifying which brain "hardware" is involved, they possess fundamental limitations in explaining the core dynamic process of NSSI. Existing research is mostly correlational and static. It effectively answers the "what" question—which brain regions are activated—but struggles to elucidate the more critical questions of "why" and "how" an internal state of emotional accumulation can dynamically and cascadingly transform into an overt, concrete act of self-injury. What is missing is a Process Model capable of describing the flow of information, the conversion of energy, and the transition of states. Merely identifying the relevant brain regions is akin to knowing a computer's CPU and memory without understanding the "algorithms" and "programs" running on them, thus failing to truly comprehend how the system derives a specific output (self-injury) from a given input (emotional distress).

Furthermore, we argue that attributing the occurrence of NSSI solely to an isolated, organic brain dysfunction may overlook its deeper roots as a mental phenomenon, namely the complex interplay between the socio-cultural environment and the individual's neurodevelopmental trajectory \cite{resch2021adolescent}. The brain's functional topology is not entirely predetermined by genetics but is continuously shaped by experience and environment throughout development. For instance, in certain cultural contexts, linguistic metaphors like "heartache" that link psychological pain to specific organ sensations may subliminally shape how individuals experience and express emotions, providing a cultural script for somatic expression \cite{grassmann2025somatic}. Moreover, in the context of modern society, the period of human socialization and maturation has been significantly extended, and the information environment has become increasingly complex. This may lead to a situation where adolescents, while facing unprecedented challenges, experience a relative delay in the maturation and differentiation of their emotional systems. This "developmental asynchrony" might be one of the contemporary reasons for the recent high prevalence of NSSI: when an individual's capacity for mentalization—the ability to distinguish and represent their own and others' mental states—is not fully developed, and the separation of emotional-somatic neural pathways is incomplete, intense internal conflicts lack mature channels for resolution.

Therefore, the study of NSSI cannot be confined to static, structural analysis but must incorporate a theoretical framework that accommodates dynamic processes and a developmental perspective. We need a model that can not only describe the real-time flow of energy or information between functional nodes but also embody how the "rules" governing this flow (i.e., the topology) are themselves adjusted and shaped by the individual's unique developmental background and environmental factors, ultimately forming a "vulnerable" pattern prone to somatic expression. This represents a core gap in the current research paradigm and is precisely the gap this study attempts to fill by constructing a computable functional topology network model.

\subsection{A New Theoretical Framework: The "Functional-Energetic Topology Model" of NSSI}

To address the gaps in existing research concerning dynamic processes and developmental perspectives, this study introduces a novel theoretical framework: the "Functional-Energetic Topology Model". This model is not a direct mapping of the brain's anatomical structure but rather a functional and abstract computational framework inspired by phenomenology, psychodynamics, and Gestalt psychology, designed to simulate the internal psychodynamic processes leading to NSSI behaviors.

\subsubsection{Core Concept: Emotion as a Dynamic Process and "Blocked Energy-Level Transition"}
The core concept of this model draws from the long-standing tradition in psychology of dynamizing emotional processes, dating back to Freud's "libido" \cite{peters1956freud}, and integrates it with Eastern philosophical understandings of systemic imbalance \cite{watts2017psychotherapy}. We view the accumulation and release of emotions as a dynamic process rather than a static affective state. Within this framework, we posit that the primary driver of NSSI is a "blocked energy-level transition"—that is, the pathway through which an internal emotional impetus flows, transforms, and reaches equilibrium within the system is obstructed. This is analogous to the concept of "qi and blood stagnation" in Traditional Chinese Medicine \cite{wang2011basic}, where energy accumulates and erupts in an abnormal or destructive manner when normal channels of release are blocked. Consequently, NSSI is regarded as the overt behavioral manifestation of this internal systemic imbalance.

\subsubsection{Topological Structure: A Phenomenological Modeling of Functional Relationships}
\begin{figure}[htbp]
    \centering
    \includegraphics[width=1\linewidth]{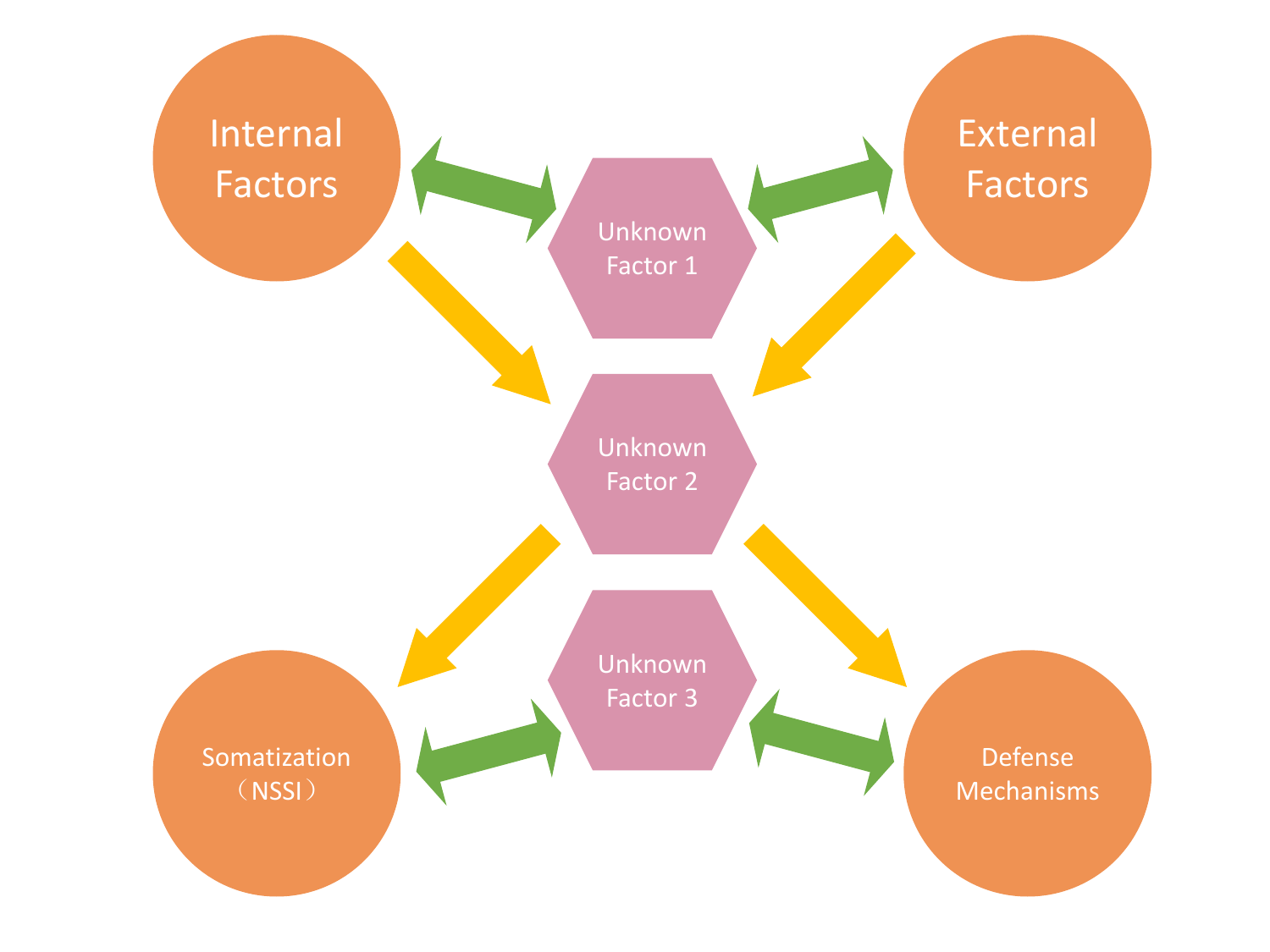}
    \caption{Schematic diagram of the proposed graph structure to explain the process of NSSI.}
    \label{fig:model_diagram}
\end{figure}

To concretize the aforementioned dynamic process, we constructed a functional topology network model. This network defines the key functional nodes for processing emotional impetus and their interrelations, simulating the complete pathway from emotional arousal to final behavioral output. Just as a geographical model explains regional climate through mountains and rivers, our topological model explains why emotional energy might "stagnate" or flow in a specific direction through the layout and connections of its nodes. The network comprises the following core components:

\begin{itemize}
    \item \textbf{Emotional Arousal Nodes:} Drawing from the phenomenological distinction between internal and external sources of intentionality \cite{wilson2003intentionality}, we divide the origin of emotions into an \textbf{Endogenous Factor} node and an \textbf{Exogenous Factor} node. These represent the emotional impetus generated by internal factors like individual temperament and genetics, and by external events such as environmental stressors, respectively.
    \item \textbf{Outcome Nodes:} The emotional impetus ultimately has two primary output pathways. One is through the \textbf{Defense Mechanism} node, where it is effectively regulated and resolved \cite{xia2023correlation}, representing healthy psychological adaptation. The other is accumulation in the \textbf{Somatization} node \cite{raffagnato2020using}; when its energy exceeds a specific threshold, it may trigger NSSI behaviors.
    \item \textbf{Transformation and Regulation Nodes:} We recognize that the process from arousal to output is not linear but is modulated by the individual's unique "function". These modulating factors are exceedingly complex, encompassing genetics, culture, developmental experiences, and more. To represent this process in the model, we have placed three \textbf{"Other"} nodes as intermediaries along the critical flow paths. These nodes do not represent any specific psychological function but are intended to simulate the individually variant adjustments, transformations, and dissipations that inevitably occur as energy flows. They collectively form two interconnected loops, creating a topology akin to the number "8", which endows the model with necessary nonlinear dynamic properties, as illustrated in Fig.~\ref{fig:model_diagram}.
\end{itemize}

\subsubsection{Core Hypotheses: Structural Ambiguity and "LOD Collapse"}
The most fundamental hypothesis of this research is rooted in the holism of Gestalt psychology \cite{stadler1994gestalt} and reflections on the nature of subjective phenomena. We postulate that the functional topological structure itself does not uniquely determine the mental phenomenon. Akin to a superposition state in quantum mechanics, even if two brains possess identical neural topologies and signal states, their corresponding subjective conscious experiences could be entirely different. This implies that even with vast amounts of high-precision brain signal data, we might still be unable to predict the inevitable occurrence of NSSI because we lack the a priori interpretive framework that endows the structure with "meaning" \cite{bealer1999theory}.

To address this, our study introduces the Level of Detail (LOD) theory as a computational simulation of this interpretive framework \cite{reddy1997perceptually}. Originating from computer graphics, the core idea of LOD is to capture meaningful, macroscopic, holistic patterns by simplifying complex structures. In our model, LOD plays the role of the "observer": we hypothesize that through a biased, simplified "observational" method (i.e., LOD-ification), this ambiguous and infinitely detailed functional network can be "collapsed" into a macroscopic, interpretable state that correlates with a clinical phenomenon (such as NSSI). Therefore, our core hypotheses can be further elaborated as follows:
\begin{enumerate}
    \item \textbf{Structural Vulnerability Hypothesis:} Patients with NSSI possess an inherent, vulnerable functional topology. Within this structure, due to specific connection weights or nodal functional properties, emotional impetus is naturally inclined to flow towards the "Somatization" node rather than being effectively neutralized by the "Defense Mechanism" node.
    \item \textbf{LOD Observability Hypothesis:} This underlying "vulnerability" cannot be directly discovered by observing all microscopic signals but can be revealed through an LOD-guided analysis. By simplifying and focusing, we can identify the macroscopic, holistic energy flow patterns that the topological network exhibits in the NSSI state. The ultimate goal of this research is to verify, through experimental data, whether this analytical framework, based on LOD theory, can successfully identify the specific topological dynamics associated with NSSI that we have hypothesized from single-channel EEG signals \cite{tong2025perception}.
\end{enumerate}

\subsection{Research Objectives and Hypotheses}
Based on the theoretical framework described above, this study aims to translate the abstract "Functional-Energetic Topology Model" into a computational paradigm that can be tested with empirical data. As we are not experts in the field of computer science, we are committed to integrating clinical insights with cutting-edge algorithms through collaboration with AI specialists. The core objective of this research is: to conduct a preliminary computational validation of the effectiveness and interpretability of the proposed functional topology theory for NSSI by constructing a Graph Neural Network (GNN) model.

To achieve this overarching goal and systematically explore the model's performance and theoretical implications, we break down the research task into the following three specific, progressive research hypotheses:
\begin{enumerate}
    \item \textbf{Hypothesis 1: Effectiveness of State Differentiation.} As a foundational validation of the model's efficacy, we hypothesize that the constructed GNN model can, with high accuracy and reliability, differentiate between the EEG states of a single patient during an NSSI-impulsive period and a non-impulsive period (resting or emotionally stable state) from their Fp1 single-channel EEG signals. This is the first step in examining whether our model can capture meaningful neural activity patterns related to the core psychological states of NSSI.
    \item \textbf{Hypothesis 2: Potential for Cross-Subject Generalization.} While an individualized model for a single patient holds significant clinical value, a core task of scientific research is to seek universal principles. Therefore, we further hypothesize that the topological features learned by the GNN model for distinguishing NSSI states possess a degree of commonality, endowing it with the capacity for cross-subject generalization. This means a model trained on data from a subset of patients can make effective predictions on the NSSI state of a new, unseen patient.
    \item \textbf{Hypothesis 3: Interpretability of Topological Features.} Our ultimate goal is not only to predict but also to explain. We hypothesize that by analyzing the trained GNN model, we can identify key topological features significantly associated with the NSSI state. A successful validation of this hypothesis would mean our theoretical model is not just an effective "black box" but an interpretable "white box". This could manifest as the model assigning higher weights to certain "edges" (representing functional pathways) or certain "nodes" (representing functional modules) exhibiting unique activation patterns when differentiating NSSI states.
\end{enumerate}

\section{Methods}\label{sec:Methods}
\subsection{Participants}

This study recruited three adolescent patients who met the diagnostic criteria for Non-Suicidal Self-Injury (NSSI). All three participants were female. To protect their privacy, all personal information was anonymized, and the participants are referred to as Patients A, B, and C in the study.

The demographic and core clinical characteristics of the participants are as follows: Patients A and B were both 14 years old, while Patient C was 16 years old. All three participants were seeking professional psychiatric help for NSSI for the first time and had not received any form of psychiatric medication prior to their consultation. According to their self-reports and those of their guardians, all three patients had a clear history of self-injury lasting approximately three months. Their NSSI behavioral patterns were highly similar, primarily involving the use of sharp objects like small blades to create multiple, dense, superficial linear cuts on the inner forearm. At the time of clinical assessment, more than thirty old and new scars were visible on their forearms. Regarding family structure, all three participants grew up in intact nuclear families with both parents present and not divorced; Patients B and C were only children, while Patient A was the second child in her family.

The study protocol strictly adhered to the Declaration of Helsinki and was reviewed and approved by the Ethics Committee of the Inner Mongolia People's Hospital. The context of this research originated from a prior clinical project aimed at alleviating adolescent anxiety through digital therapeutics. We provided all potential participants and their legal guardians with a detailed explanation of the study's purpose and procedures, emphasizing that the EEG biofeedback training used in this research is an auxiliary method designed to help them with emotional relaxation and to enhance their self-regulation skills. All participants and their guardians signed written informed consent forms on a fully understood and voluntary basis. Participants were informed that while using the provided application for daily relaxation training, if they experienced a persistent urge to self-injure during a session, they could mark that specific record through a designated function in the software after the session. This design allowed us to naturally acquire valuable EEG data related to the NSSI-impulsive state in a manner that was ethically compliant and minimally intrusive to the participants.

\subsection{EEG Data Acquisition}
The EEG data for this study was collected within the context of a home-based digital therapeutic application designed for the participants \cite{arntz2023technologies}. Our research team optimized and iterated on a previously developed biofeedback meditation software for anxiety disorder treatment, making the relaxation and meditation process more simple and efficient for adolescents, while also adding a feature to confirm the user's current NSSI state. Through these specific software designs, we aimed to acquire neural activity data related to NSSI in an ecological and minimally intrusive manner.

\subsubsection{Acquisition Device and Parameters}
We used a commercial portable EEG acquisition device—the MindWave Mobile 2 headset from NeuroSky \cite{morshad2020analysis}. This device has a built-in TGAM (ThinkGear ASIC Module) chip and collects single-channel EEG signals via a single dry electrode at the Fp1 position of the international 10-20 system. The device's original sampling rate is 512 Hz. The TGAM chip not only outputs raw EEG waveform data but also provides real-time proprietary high-level metrics such as "Attention" and "Meditation," as well as power values for different frequency bands (e.g., Delta, Theta, Alpha, Beta, Gamma) \cite{yin2020experimental}. The complete structure of the data packets output by the device is detailed in Table~\ref{tab:byte_value}.

\begin{table}[htbp] 
    \centering
    \caption{The data structure in EEG packets sent by TGAM Bluetooth}
    \label{tab:byte_value}
    \resizebox{\columnwidth}{!}{%
    \begin{tabular}{ccl}
    \toprule
    \textbf{Byte} & \textbf{Value} & \textbf{Explanation} \\
    \midrule
    1 & 0xAA & [SYNC] \\
    2 & 0x20 & [PLENGTH] (payload length) of 32 bytes \\
    3 & 0x00 & [POOR\_SIGNAL] Quality \\
    4 & 0x83 & No poor signal detected (0/200) \\
    8 & 0x18 & [ASIC\_EEG\_POWER\_INT] \\
    9 & 0x20 & [VLENGTH] 24 bytes \\
    10 & 0x00 & (1/3) Begin Delta bytes \\
    11 & 0x00 & (2/3) \\
    12 & 0x94 & (3/3) End Delta bytes \\
    13 & 0x00 & (1/3) Begin Theta bytes \\
    14 & 0x00 & (2/3) \\
    15 & 0x42 & (3/3) End Theta bytes \\
    16 & 0x00 & (1/3) Begin Low-alpha bytes \\
    17 & 0x00 & (2/3) \\
    18 & 0x0B & (3/3) End Low-alpha bytes \\
    19 & 0x00 & (1/3) Begin High-alpha bytes \\
    20 & 0x00 & (2/3) \\
    21 & 0x64 & (3/3) End High-alpha bytes \\
    22 & 0x00 & (1/3) Begin Low-beta bytes \\
    23 & 0x00 & (2/3) \\
    24 & 0x4D & (3/3) End Low-beta bytes \\
    25 & 0x00 & (1/3) Begin High-beta bytes \\
    26 & 0x00 & (2/3) \\
    27 & 0x3D & (3/3) End High-beta bytes \\
    28 & 0x00 & (1/3) Begin Low-gamma bytes \\
    29 & 0x00 & (2/3) \\
    30 & 0x07 & (3/3) End Low-gamma bytes \\
    31 & 0x00 & (1/3) Begin Mid-gamma bytes \\
    32 & 0x00 & (2/3) \\
    33 & 0x05 & (3/3) End Mid-gamma bytes \\
    34 & 0x04 & [ATTENTION] eSense \\
    35 & 0x0D & eSense Attention level of 13 \\
    36 & 0x05 & [MEDITATION] eSense \\
    37 & 0x3D & eSense Meditation level of 61 \\
    38 & 0x34 & [CHKSUM] (1’s comp inverse of 8-bit \\ 
       &      & Payload sum of 0xCB) \\
    \bottomrule
    \end{tabular}%
    }
\end{table}

\subsubsection{Data Collection Paradigm and Procedure}
All data were collected using an application developed by our team, installed on the participants' Android phones or tablets. According to the research protocol, we recommended that participants use this application as a daily emotion regulation tool, conducting three ten-minute relaxation sessions per day: in the morning, at noon, and in the evening.

\begin{figure*}[ht]
    \centering
    \includegraphics[width=1\textwidth]{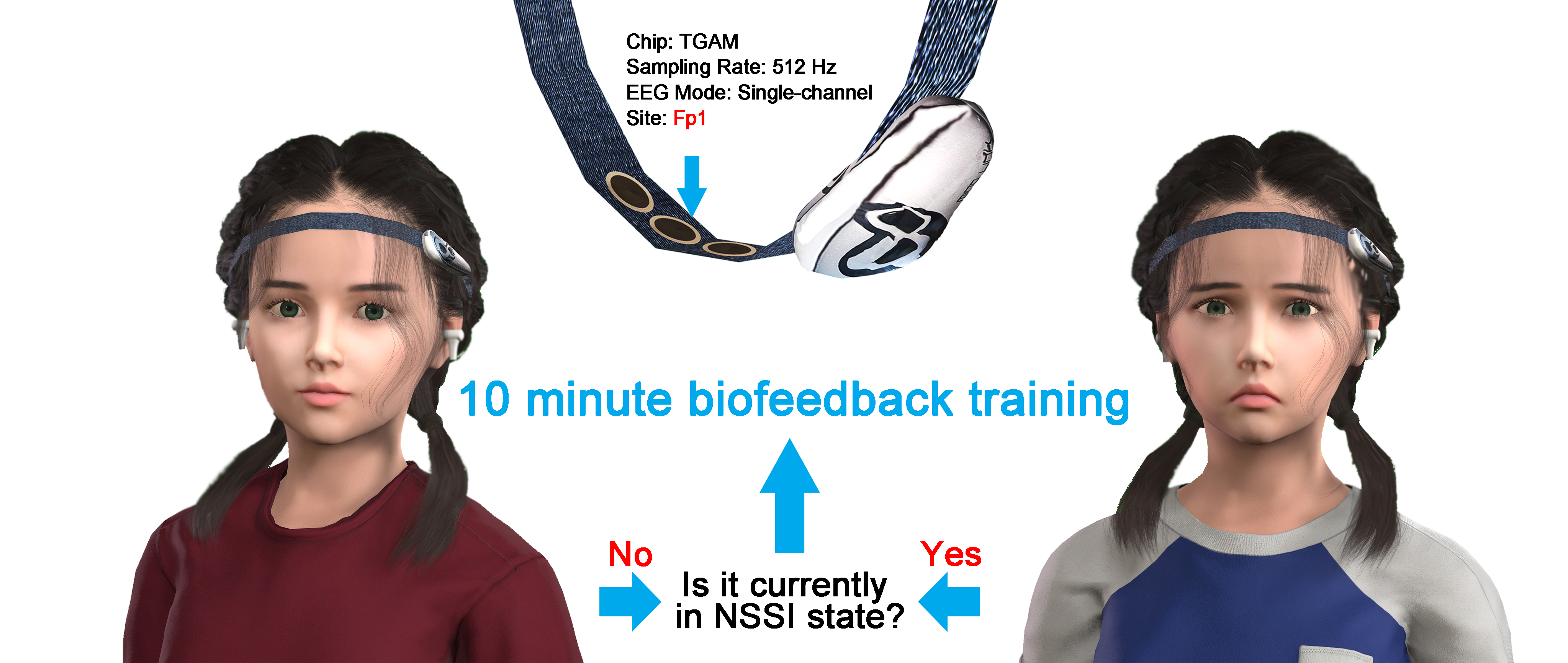}
    \caption{Schematic diagram of the process where patients wear the EEG headset to complete music-guided meditation and relaxation in different states.}
    \label{fig:data_collection_procedure}
\end{figure*}

The training procedure was as follows:
\begin{enumerate}
    \item \textbf{Guidance and Feedback:} After the participant put on the EEG headset and connected it to the application via Bluetooth, the program would start playing a fixed background music with a clear rhythm. Simultaneously, an "energy ball" would be displayed on the screen as a visual interface for real-time biofeedback. The size or brightness of this energy ball would change according to the "Meditation" metric transmitted in real-time from the headset, thus providing the participant with intuitive feedback on their current state of relaxation and guiding them in active self-regulation.
    \item \textbf{Data Recording and Labeling:} During each 10-minute session, the application would fully record the entire stream of EEG data transmitted from the headset. After the session concluded, the program would ask the participant if they experienced a strong urge to self-injure during the session. If the participant answered "Yes," the 10-minute EEG data segment would be automatically labeled as an "NSSI state"; if they answered "No," it would be labeled as a "non-NSSI state," as shown in Fig.~\ref{fig:data_collection_procedure}.
\end{enumerate}

\subsubsection{Dataset Composition}
Through the collection paradigm described above, we gathered a total of 231 effective 10-minute EEG data segments from the three participants over a period of approximately one month. Among these, 169 segments were defined as "non-NSSI state" data, and 62 segments were "NSSI state" data. The specific data distribution contributed by each participant is detailed in Table~\ref{tab:dataset_composition}. This data forms the complete dataset for the subsequent model training and validation in this study.

\begin{table}[htbp]
    \centering
    \caption{EEG data classification statistics collected from the three patients in different states}
    \label{tab:dataset_composition}
    \resizebox{\columnwidth}{!}{%
    \begin{tabular}{lccc}
        \toprule
        \textbf{Participant} & \makecell{\textbf{Non-NSSI} \\ \textbf{Records (count)}} & \makecell{\textbf{NSSI} \\ \textbf{Records (count)}} & \makecell{\textbf{Total} \\ \textbf{Records (count)}} \\
        \midrule
        Patient A & 57 & 21 & 78 \\
        Patient B & 47 & 18 & 65 \\
        Patient C & 65 & 23 & 88 \\
        \midrule
        \textbf{Total} & \textbf{169} & \textbf{62} & \textbf{231} \\
        \bottomrule
    \end{tabular}%
    }
\end{table}

\subsection{Data Preprocessing and Feature Engineering}
To transform the raw, continuous EEG signals collected from the Fp1 electrode into clean, structured data suitable for Graph Neural Network (GNN) model analysis, we designed and executed the following series of preprocessing and feature engineering steps.

\subsubsection{Noise Reduction and Artifact Removal}
We recognize that EEG signals from the Fp1 position are highly susceptible to contamination from physiological artifacts such as eye movements (EOG), blinks, and frontal muscle activity (EMG) \cite{wu2020attention}. Additionally, as a consumer-grade wireless device, its signal may also contain environmental noise \cite{sabio2024scoping}. To enhance signal quality, we implemented the following key noise reduction steps:
\begin{enumerate}
    \item \textbf{Band-pass Filtering:} We first applied a 1-50 Hz band-pass filter to the raw data \cite{ronca2024optimizing}. This step was aimed at removing low-frequency skin potential drifts and high-frequency power-line noise (50/60 Hz) as well as muscle activity artifacts, thereby preserving the core EEG frequency band information crucial for our research.
    \item \textbf{Artifact Removal:} Subsequently, we employed Independent Component Analysis (ICA) to further separate and remove residual physiological artifacts \cite{mannan2018identification}. The ICA algorithm decomposes the mixed EEG signals into multiple independent source components, allowing us to identify and remove those components that exhibit typical eye movement or blink patterns on a topographic map, ultimately reconstructing a purer EEG dataset.
\end{enumerate}

\subsubsection{Signal Downsampling}
Considering that the MindWave Mobile 2 headset used in this study transmits data wirelessly via Bluetooth, we downsampled the denoised 512 Hz data to ensure the stability of the data stream during long-term recordings and to minimize potential data packet loss due to signal transmission instability. We adopted a method of sampling every other data point, reducing the sampling rate to 256 Hz. This pragmatic step was a decision based on our long-term experience with this type of consumer-grade device, aimed at ensuring the integrity and reliability of the data fed into the subsequent analysis pipeline.

\subsubsection{Data Segmentation and Sample Construction}
Following our core idea of analyzing EEG states from a dynamic perspective, we segmented each preprocessed, continuous 10-minute (600-second) EEG recording into 600 consecutive, non-overlapping, 1-second fixed-duration windows (Epochs). Through this segmentation, we effectively transformed a lengthy time-series data stream into a sequence of 600 independent samples. Each 1-second window is considered a "snapshot" reflecting the brain's functional state at that moment and serves as the basic unit for classification and prediction by the GNN model.

\subsubsection{Frequency-Domain Feature Extraction}
For each 1-second data window, we performed a frequency-domain analysis to extract core features that characterize its neural activity. Specifically, we calculated the Power Spectral Density (PSD) within the window using the Fast Fourier Transform (FFT) \cite{heckbert1995fourier} and Welch's method \cite{solomon1991psd}, and extracted the power of five classic EEG frequency bands, which are widely associated with various cognitive and emotional states, as features: Delta band (1-4 Hz), Theta band (4-8 Hz), Alpha band (8-13 Hz), Beta band (13-30 Hz), and Gamma band (30-50 Hz).

It is worth noting that we intentionally selected these fundamental and widely accepted frequency-domain features, rather than using the device's proprietary high-level metrics like "Attention" and "Meditation". This is because the update frequency and stability of the latter during data transmission were insufficient to meet the second-level precision required for our model's dynamic analysis. Ultimately, each 1-second data window was converted into a five-dimensional feature vector, which serves as the raw input for our proposed functional topology network model.

\subsection{GNN Model Construction}
The core of this research lies in transforming our proposed "Functional-Energetic Topology Model" into a trainable, quantitative computational model. We chose the Graph Neural Network (GNN) \cite{scarselli2008graph} as the ideal tool to achieve this goal, as it is naturally adept at handling relational data between nodes and can simulate the flow and interaction of information within our theoretical framework. The construction, training, and validation of the model were all completed with the collaboration and assistance of AI specialists.

\subsubsection{Graph Definition}
We defined a fixed directed graph with 7 nodes and multiple edges to simulate the core psychodynamic pathways related to NSSI. This structure is a direct projection of our theoretical model, as shown in Fig.~\ref{fig:model_diagram}, and its components are as follows:
\begin{itemize}
    \item \textbf{Nodes:} The graph contains 7 functional nodes, representing:
    \begin{itemize}
        \item Two emotional arousal sources: the \textbf{Endogenous Factor} node and the \textbf{Exogenous Factor} node.
        \item Two emotional outcome pathways: the \textbf{Defense Mechanism} node and the \textbf{Somatization} node.
        \item Three transformation and regulation intermediaries: \textbf{Other Nodes 1, 2, and 3}.
    \end{itemize}
    \item \textbf{Edges:} The connections between nodes (i.e., edges) are predefined to represent our hypothesized directions of energy flow. For example, there are unidirectional edges from the "Endogenous/Exogenous" nodes to the "Defense/Somatization" nodes, as well as bidirectional edges between the "Endogenous" and "Exogenous" nodes. Together, these form the topological backbone of the model.
\end{itemize}

\subsubsection{Feature Mapping}
In this study, the observed values from the Fp1 single-channel EEG are considered an "external probe" of the macroscopic state of the entire functional topology network. Therefore, we use the five-dimensional frequency-domain feature vector extracted from each 1-second data window as the initial input that drives the dynamics of the entire network.

In practice, this five-dimensional feature vector is simultaneously assigned to all 7 nodes in the graph as their \textbf{Initial Node Features} for each time step (i.e., each 1-second window). We believe that traditional EEG analysis is akin to processing the acoustic signals of an unknown language, where the model can only find statistical patterns in the raw waveforms. The topological structure proposed in this research, however, provides the model with an \textbf{\textit{a priori} grammatical framework}. When the raw neural signals flow and interact within this predefined grammatical framework, their intrinsic "meaning" related to the clinical state can be more effectively decoded by the model.

\subsubsection{GNN Architecture}
\begin{figure*}[ht]
    \centering
    \includegraphics[width=1\textwidth]{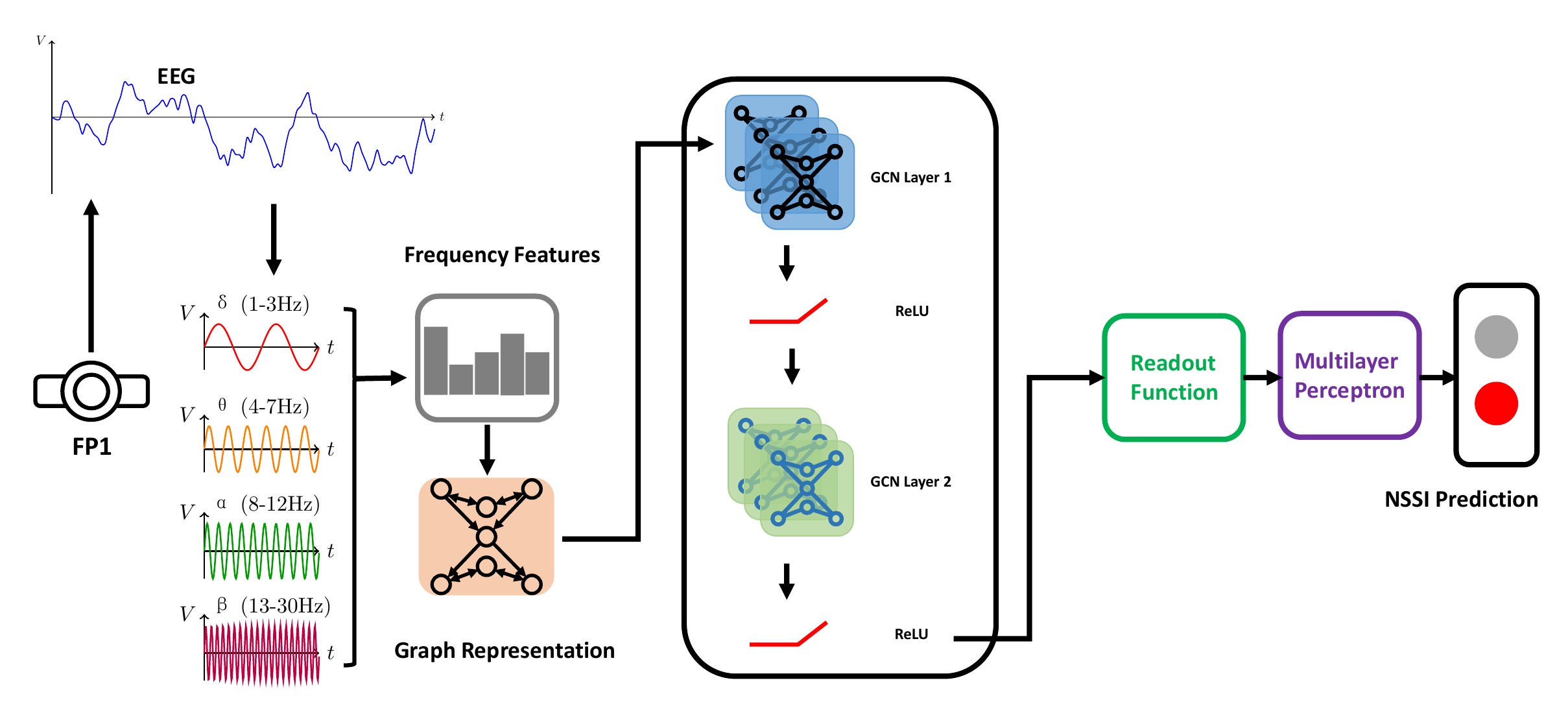}
    \caption{Schematic diagram of the GNN model's construction, training, and validation process.}
    \label{fig:model_architecture}
\end{figure*}

We constructed a model based on the Graph Convolutional Network (GCN) \cite{zhang2019graph}, with an architecture designed to simulate the propagation and integration of information within the aforementioned topological structure. The specific layers and parameter design of the model are as follows, and a schematic is shown in Fig.~\ref{fig:model_architecture}:
\begin{enumerate}
    \item \textbf{Input Layer:} The model's input is the graph defined for each 1-second window. The node feature matrix for each graph has a dimension of (7, 5), representing 7 nodes, each with a 5-dimensional feature vector.
    \item \textbf{Graph Convolutional Layers:} The model contains two consecutive GCN layers. Each GCN layer aggregates information from neighboring nodes based on the graph's adjacency structure to update the feature representation of the central node. This process computationally simulates the one-step propagation and interaction of "energy" or "information" along the predefined "edges".
    \item \textbf{Hidden Dimension:} To balance computational efficiency with model performance in the initial exploratory phase, we set the hidden dimension of the GCN layers to 8. This means that after the first GCN layer, the feature vector of each node is mapped from 5 dimensions to 8, allowing it to capture richer combinatorial features.
    \item \textbf{Activation Function:} After each GCN layer, we applied the Rectified Linear Unit (ReLU) as the activation function. ReLU introduces non-linearity into the model, enabling it to learn more complex relationships between input and output that go beyond simple linear mappings.
    \item \textbf{Output Layer:} After propagation through two graph convolutional layers, we perform Global Average Pooling on the final feature vectors of all 7 nodes to obtain a single vector that represents the state of the entire graph. This vector is then passed through a Fully Connected Layer, which compresses its dimension to 1. Finally, a Sigmoid activation function transforms this value into a probability score between 0 and 1, representing the likelihood of an NSSI state occurring in the current 1-second window.
\end{enumerate}

\subsection{Experimental Design and Model Evaluation}
To systematically evaluate the performance of our constructed Graph Neural Network (GNN) model and to validate the three core hypotheses proposed in this study, we designed a comprehensive experimental protocol that includes both intra-subject validation and cross-subject generalization assessment.

\subsubsection{Supervised Learning Task Definition and Label Assignment}
The core task of this research is defined as a binary classification supervised learning problem. We treat each preprocessed 1-second EEG data window as an independent sample. Based on the participants' subjective reports at the time of data collection, we assigned labels to these samples as follows:
\begin{itemize}
    \item \textbf{Label "1" (Positive Sample):} Assigned to EEG windows during which the participant reported being in an "NSSI state".
    \item \textbf{Label "0" (Negative Sample):} Assigned to EEG windows during which the participant reported being in a "non-NSSI state".
\end{itemize}
The training objective of the model is to learn an accurate mapping from the input graph data (representing the EEG state) to its corresponding label (representing the clinical state).

\subsubsection{Intra-subject Validation}
This validation scheme aims to test Hypothesis 1, i.e., the model's effectiveness in distinguishing between different states within a single patient. We conducted separate training and evaluation for each participant's (A, B, and C) dataset:
\begin{enumerate}
    \item \textbf{Data Split:} For each patient's entire set of data windows, we performed a random split, allocating 80\% as the training set and 20\% as the test set.
    \item \textbf{Independent Training:} We trained three separate GNN models, one for each patient. Each model was trained using only the training set data of its corresponding patient.
    \item \textbf{Performance Evaluation:} After training was complete, the performance of each model was evaluated on its respective independent test set. This strict separation ensures the objectivity of the evaluation results, preventing the model from being tested on data it had already "seen".
\end{enumerate}

\subsubsection{Inter-subject Validation}
To test Hypothesis 2, i.e., whether the model possesses the potential for cross-subject generalization, we employed a Leave-One-Subject-Out Cross-Validation (LOSOCV) strategy \cite{pauli2021balanced}. This strategy is the gold standard for assessing model generalization on small sample sizes. The specific procedure was as follows:
\begin{enumerate}
    \item \textbf{Rotational Training and Testing:} We conducted three independent rounds of experiments.
    \begin{itemize}
        \item \textbf{Round 1:} Used all data from Patients A and B as the training set and tested on all data from Patient C.
        \item \textbf{Round 2:} Used all data from Patients A and C as the training set and tested on all data from Patient B.
        \item \textbf{Round 3:} Used all data from Patients B and C as the training set and tested on all data from Patient A.
    \end{itemize}
    \item \textbf{Comprehensive Evaluation:} Finally, we averaged the performance metrics from the test sets across the three rounds to obtain a comprehensive assessment of the model's cross-subject generalization capability.
\end{enumerate}

\subsubsection{Model Training Details and Evaluation Metrics}
\begin{itemize}
    \item \textbf{Loss Function:} Given that this is a binary classification task, we chose Binary Cross-Entropy Loss as the objective function for model training. This function measures the discrepancy between the model's predicted probabilities and the true labels (0 or 1).
    \item \textbf{Optimizer:} We used the Adam optimizer to update the model's network weights, with an initial learning rate set to 0.001. Adam is an efficient and adaptive optimization algorithm widely used in deep learning tasks.
    \item \textbf{Evaluation Metrics:} To comprehensively and objectively evaluate the model's performance, we used the following series of standard metrics:
    \begin{itemize}
        \item \textbf{Accuracy:} The proportion of samples correctly classified by the model out of the total number of samples.
        \item \textbf{Precision:} Among all samples predicted as "NSSI state," the proportion that are truly "NSSI state."
        \item \textbf{Recall / Sensitivity:} Among all true "NSSI state" samples, the proportion that were successfully predicted by the model.
        \item \textbf{F1-Score:} The harmonic mean of Precision and Recall, serving as a key indicator of a model's overall performance, especially in cases of imbalanced data.
        \item \textbf{Confusion Matrix:} Provides an intuitive visualization of the model's prediction performance across different classes.
        \item \textbf{Area Under the ROC Curve (AUC):} Measures the model's overall ability to distinguish between classes across all possible thresholds. A value closer to 1 indicates better model performance.
    \end{itemize}
\end{itemize}

\section{Results}\label{sec:Results}
This chapter systematically presents our experimental findings to sequentially validate the three core hypotheses proposed in the introduction. We will first describe the characteristics of the final dataset used for model construction, then report the model's performance on both intra-subject and cross-subject generalization tasks, and finally, through explainability analysis, reveal the key topological features associated with the NSSI state.

\subsection{Dataset Description}
Following the data preprocessing and feature engineering pipeline, we ultimately constructed a complete dataset for the training and validation of the GNN model. This dataset is composed of 1-second EEG window samples from the three NSSI patients.

As detailed in Table~\ref{tab:dataset_overview}, the final dataset for analysis comprises a total of \textbf{138,600} 1-second window samples (from 231 records $\times$ 600 seconds/record). Among these, \textbf{37,200} windows were labeled as "NSSI state" (positive samples), accounting for \textbf{26.8\%} of the total samples, while \textbf{101,400} windows were labeled as "non-NSSI state" (negative samples), making up \textbf{73.2\%} of the total. The dataset exhibits a certain degree of class imbalance, with a significantly larger number of negative samples than positive ones. In our subsequent model evaluation, we will pay special attention to metrics that are more robust to imbalanced data, such as the F1-score and AUC, to ensure an objective assessment of the model's performance.

\begin{table}[htbp]
    \centering
    \caption{Overview of the final dataset used for modeling}
    \label{tab:dataset_overview}
    \resizebox{\columnwidth}{!}{
    \begin{tabular}{lccccc}
        \toprule
        \textbf{Data Category} & \textbf{Patient A} & \textbf{Patient B} & \textbf{Patient C} & \textbf{Total} & \textbf{Proportion} \\
        \midrule
        \makecell[l]{NSSI State Windows\\(Positive Samples)} & 12,600 & 10,800 & 13,800 & \textbf{37,200} & 26.8\% \\
        \makecell[l]{Non-NSSI State\\Windows\\(Negative Samples)} & 34,200 & 28,200 & 39,000 & \textbf{101,400} & 73.2\% \\
        \midrule
        \textbf{Total Windows} & \textbf{46,800} & \textbf{39,000} & \textbf{52,800} & \textbf{138,600} & \textbf{100\%} \\
        \hline
        \makecell[l]{Original Records\\(10 min/record)} & 78 & 65 & 88 & \textbf{231} & -- \\
        \bottomrule
    \end{tabular}
    }
\end{table}

\subsection{Intra-subject Predictive Performance}
To test our first hypothesis—that the model can effectively distinguish between NSSI and non-NSSI states within an individual—we independently trained and evaluated the GNN model for each of the three participants. The experiment strictly followed the intra-subject validation protocol described in Section 2.5.2, where each patient's data was split into an 80\% training set and a 20\% test set.

The predictive performance of the model is summarized in detail in Table~\ref{tab:intra_subject_performance}. The results show that the model demonstrated excellent classification performance for all three participants. Specifically, the model achieved overall accuracy rates of 86.3\%, 84.8\%, and 85.9\% on the independent test sets for Patients A, B, and C, respectively.

Beyond accuracy, the model also showed strong performance on other key metrics. The F1-score, a crucial indicator of a model's comprehensive performance on imbalanced data, averaged around 0.763 across the three patients, indicating a good balance between precision and recall. Furthermore, the AUC values, which measure the model's overall discriminative ability, all exceeded 0.90. This robustly indicates that our GNN model can, to a certain extent, reliably and effectively identify neural activity patterns associated with NSSI impulses from a single individual's single-channel EEG signal. These findings provide solid empirical support for our first hypothesis.

\begin{figure}[htbp]
    \centering
    \includegraphics[width=\columnwidth]{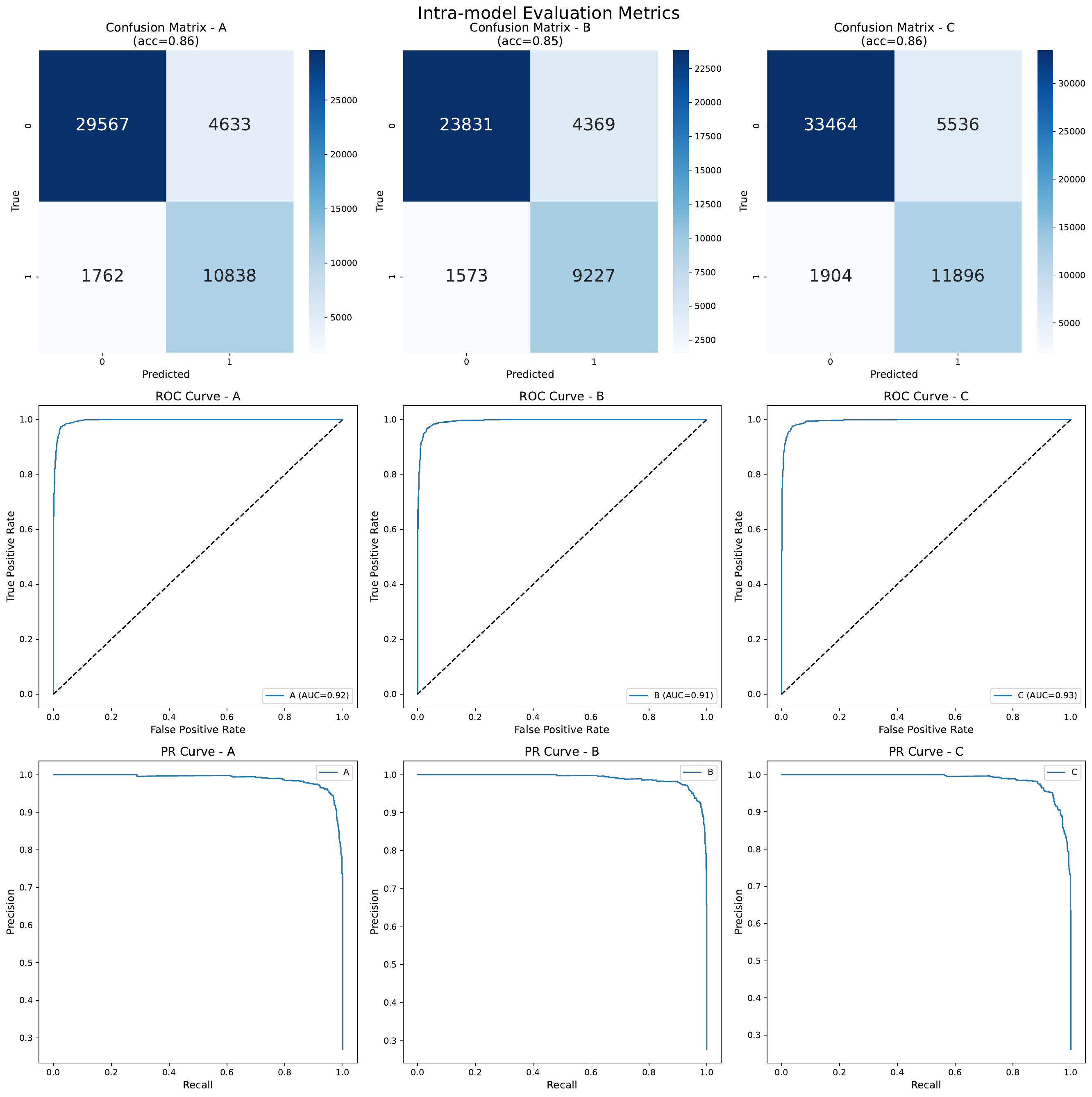}
    \caption{Confusion matrices and ROC/PR curves for intra-subject prediction.}
    \label{fig:intra_subject_curves}
\end{figure}

To provide a more intuitive illustration of the model's classification details, we present the confusion matrices and ROC curves for the test sets of the three patients in Fig.~\ref{fig:intra_subject_curves}. As can be clearly seen from the figure, the model performs exceptionally well in correctly identifying both 'NSSI states' (True Positives) and 'non-NSSI states' (True Negatives), while the proportions of confusion between the two (False Positives and False Negatives) are relatively low.

\begin{table}[htbp]
    \centering
    \caption{Intra-subject predictive performance evaluation results}
    \label{tab:intra_subject_performance}
    \resizebox{\columnwidth}{!}{%
    \begin{tabular}{lccccc}
        \toprule
        \textbf{Participant} & \textbf{Accuracy} & \textbf{Precision} & \textbf{Recall} & \textbf{F1-Score} & \textbf{AUC} \\
        \midrule
        Patient A & 0.863 & 0.701 & 0.860 & 0.772 & 0.92 \\
        Patient B & 0.848 & 0.679 & 0.854 & 0.756 & 0.91 \\
        Patient C & 0.859 & 0.682 & 0.862 & 0.762 & 0.93 \\
        \bottomrule
    \end{tabular}%
    }
\end{table}

\subsection{Cross-subject Generalization Performance}
Having validated the model's strong intra-subject predictive capabilities, we proceeded to test the more critical Hypothesis 2—whether our GNN model, based on functional topology theory, captures some common, cross-subject transferable features in the neurodynamics of NSSI. To this end, we strictly implemented the Leave-One-Subject-Out Cross-Validation (LOSOCV) scheme defined in Section 2.5.3.

The results of this scheme directly reflect the model's predictive performance when facing a completely new, unseen patient. As shown in Table~\ref{tab:cross_subject_performance}, the average results of the three cross-validation rounds indicate that the model achieved an overall accuracy of 73.7\% in the cross-subject prediction task. Although this value is, as expected, lower than that of the intra-subject models, we believe this result is almost akin to biased random guessing, or it may suggest that the model relies on "memorizing" specific patients' EEG patterns for its intra-subject predictions.

\begin{figure}[htbp]
    \centering
    \includegraphics[width=\columnwidth]{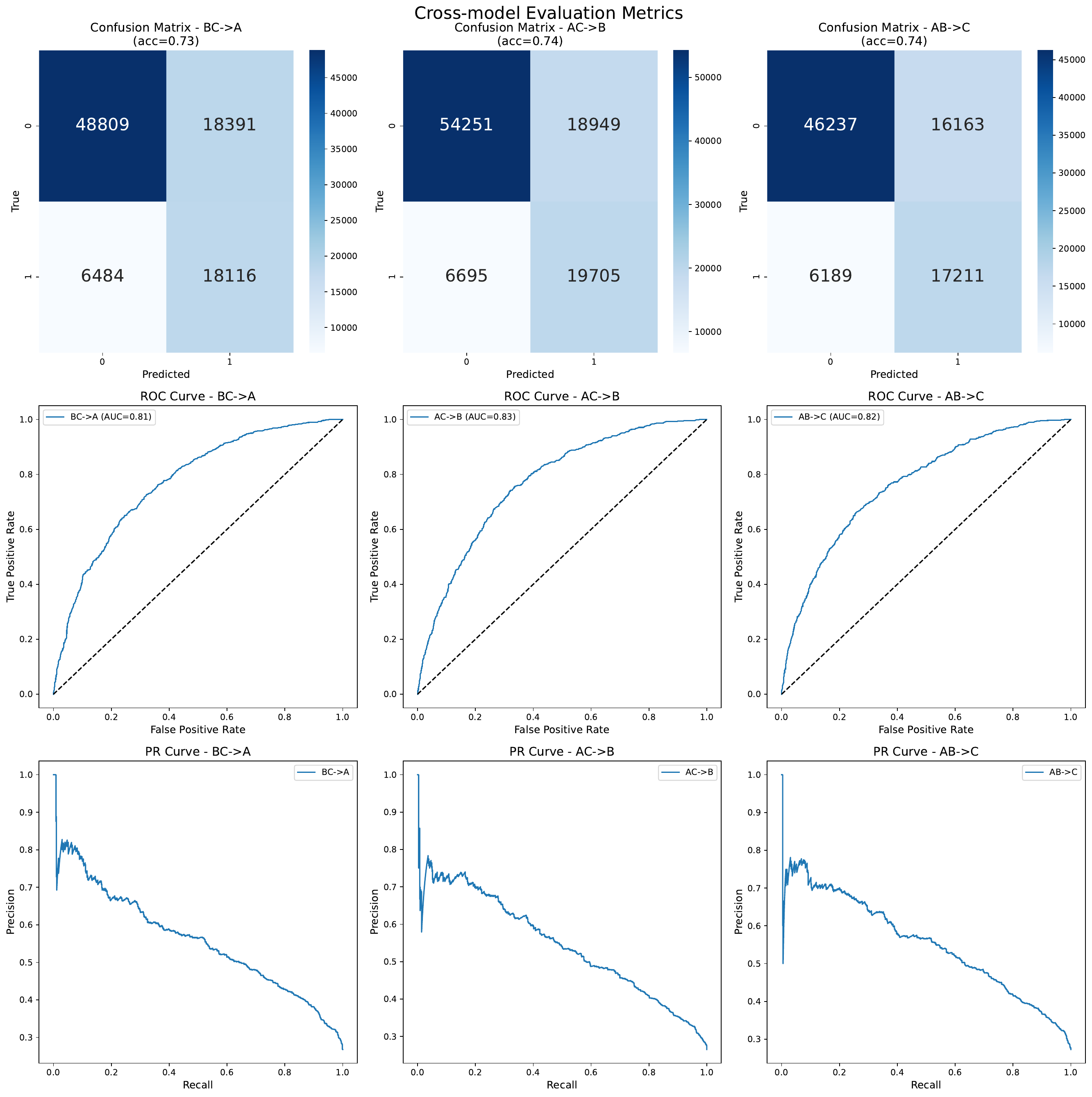}
    \caption{Confusion matrices and ROC/PR curves for the cross-subject generalization test sets.}
    \label{fig:cross_subject_curves}
\end{figure}

However, it is noteworthy that the model achieved scores of 0.602 and 0.82 on the F1-score and AUC metrics, respectively, which are more robust to class imbalance. This outcome is closely linked to the model's high recall rate. We speculate that this might indicate the model, trained under the GNN framework, has the potential to make meaningful and clinically valuable predictions for future NSSI states based on the "vulnerable topology" patterns learned from other patients. Therefore, these findings provide preliminary, though not yet solid, evidence for our second hypothesis—that the model possesses cross-subject generalization potential. They also suggest that our proposed functional topology model may have the opportunity to touch upon some common neural mechanisms underlying the NSSI phenomenon through continuous improvement and optimization. The confusion matrices and ROC curves for the cross-subject generalization test sets are presented in Fig.~\ref{fig:cross_subject_curves}.

\begin{table}[htbp]
    \centering
    \caption{Cross-subject generalization performance evaluation results (LOSOCV)}
    \label{tab:cross_subject_performance}
    \resizebox{1\columnwidth}{!}{%
    \begin{tabular}{lccccc} 
        \toprule
        \makecell[l]{Test Subject \\ \textit{(Training Set)}} & \textbf{Accuracy} & \textbf{Precision} & \textbf{Recall} & \textbf{F1-Score} & \textbf{AUC} \\
        \midrule
        \makecell[l]{Patient A \\ \textit{(B \& C)}} & 0.729 & 0.496 & 0.736 & 0.593 & 0.81 \\
        \makecell[l]{Patient B \\ \textit{(A \& C)}} & 0.743 & 0.510 & 0.746 & 0.606 & 0.83 \\
        \makecell[l]{Patient C \\ \textit{(A \& B)}} & 0.739 & 0.516 & 0.736 & 0.606 & 0.82 \\
        \bottomrule
    \end{tabular}%
    }
\end{table}

\subsection{Explainability Analysis of the Model: Key Topological Features of NSSI}
To delve into the internal decision-making logic that allows the model to distinguish between the two states, we conducted a GNNExplainer-based explainability analysis for both the NSSI and non-NSSI states \cite{ying2019gnnexplainer}. By precisely comparing the importance weights assigned by the model to various functional pathways under each condition, we were able to uncover a previously undiscovered dynamic transition mechanism related to the onset of NSSI.

First, the analysis revealed a constant background feature of this patient cohort: \textbf{high-intensity internal processing and ideation}. In both states, the pathway from the "Internal Factors" node to the "Unknown Factor 1" node maintained an extremely high weight (Non-NSSI: 0.45, NSSI: 0.46), indicating that continuous and intense internal psychological activity is a consistent baseline state.

\begin{figure*}[ht]
    \centering
    \includegraphics[width=1\textwidth]{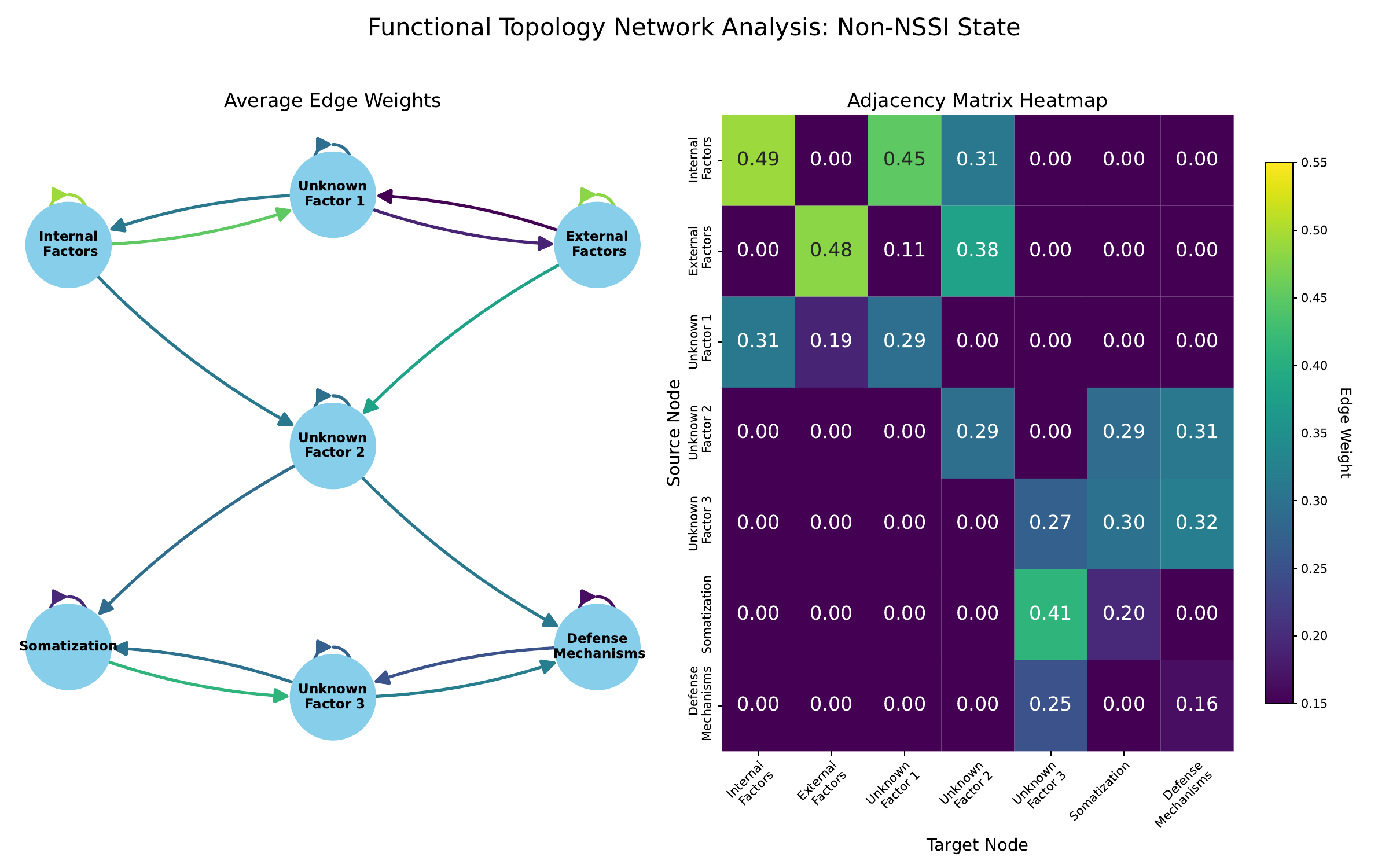}
    \caption{Functional topology network average edge weights and adjacency matrix heatmap for the Non-NSSI state.}
    \label{fig:non_nssi_weights}
\end{figure*}

\begin{figure*}[ht]
    \centering
    \includegraphics[width=1\textwidth]{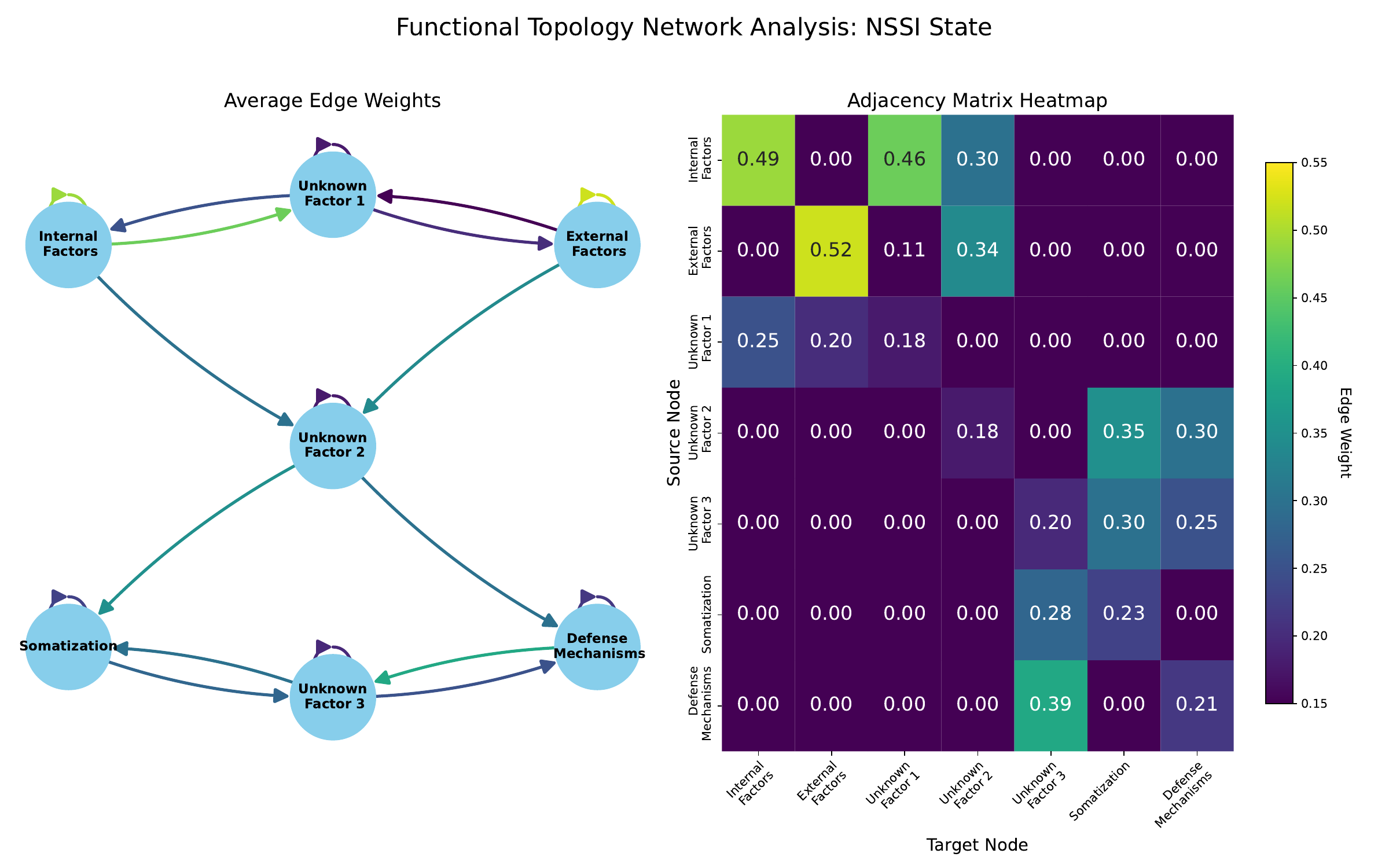}
    \caption{Functional topology network average edge weights and adjacency matrix heatmap for the NSSI state.}
    \label{fig:nssi_weights}
\end{figure*}

The most critical finding of this study lies in the identification of a \textbf{key feedback regulatory loop}, operating around the "Unknown Factor 3" node, that undergoes a \textbf{directional reversal} between the two states.
\begin{enumerate}
    \item \textbf{In the Non-NSSI State (Functional Regulatory Mode):} As shown in Fig.~\ref{fig:non_nssi_weights}, we observe a healthy "self-correction" feedback loop. The weight of the connection from the "Somatization" node to the "Unknown Factor 3" node is very high (0.41), and subsequently, the "Unknown Factor 3" node effectively transmits this information to the "Defense Mechanisms" node (weight 0.32). This clearly depicts a functional regulatory process: when the body experiences negative sensations (somatization), the system can effectively capture this signal and guide it through an intermediary (Unknown Factor 3) for processing and resolution by the defense mechanisms.
    \item \textbf{In the NSSI State (Decompensated Reversal Mode):} As shown in Fig.~\ref{fig:nssi_weights}, the aforementioned healthy loop undergoes a critical reversal. The weight of the key input pathway from "Somatization" to "Unknown Factor 3" drops sharply to 0.28. Concurrently, a reverse pathway, from "Defense Mechanisms" to "Unknown Factor 3," becomes abnormally strengthened to 0.39.
\end{enumerate}

This "feedback loop reversal" is key to explaining the trigger of the NSSI state. It implies that in the NSSI state, the system not only loses its ability to self-correct based on somatic sensations, but its own defense mechanism seems to fall into a state of "ineffective idling" or "generating interfering signals," becoming a source of maladaptive information input to the system.

Furthermore, we also observed an auxiliary change: in the NSSI state, the pathway weight from the "Unknown Factor 2" node to the "Somatization" node was enhanced (from 0.29 to 0.35), which further exacerbated the system's tendency toward somatic expression.

In summary, the model's explainability analysis provides us with a novel, data-driven dynamic model of NSSI. Its core mechanism is not a simple change in pathway strength, but rather the \textbf{dysfunction and directional reversal of a key feedback regulatory loop responsible for processing somatic sensations}, which ultimately leads to the functional dysregulation of the entire emotion regulation system and the emergence of maladaptive behaviors.

\section{Discussion}\label{sec:Discussion}

\subsection{Main Findings Summary}
This study aimed to develop and validate an innovative computational model based on a functional topology theory to uncover the neurodynamic mechanisms of NSSI from single-channel EEG signals. Through analyzing data from three adolescent patients, we have arrived at three primary findings. First, we successfully constructed a computational model capable of effectively distinguishing NSSI states, demonstrating excellent performance in intra-subject prediction tasks (average accuracy > 85\%). Second, the model exhibited preliminary but statistically significant cross-subject generalization potential, achieving an accuracy ($\approx$ 73.6\%) that is considerably better than random chance. Finally, and most insightfully, our explainability analysis revealed a highly specific dynamic dysregulation mechanism for the onset of the NSSI state. The results indicate that the trigger for NSSI is not the simple activation of a pathological pathway, but rather a complex systemic event, the core of which is the functional reversal of a key feedback regulatory loop.

\subsection{Empirical Support for the Theoretical Model}
The most central contribution of this research lies not only in building an effective predictive model but also in how the model's explainability analysis provides direct and profound empirical support for our initially proposed "Functional-Energetic Topology Model," while also unexpectedly refining and deepening it. Our findings clearly reveal that NSSI is not a simple linear process but a complex dynamic system failure event, the core of which is the \textbf{functional reversal of a key feedback regulatory loop}.

Our initial theoretical hypothesis leaned towards the idea that NSSI originated from the activation of a "vulnerable pathway" leading from emotional arousal to somatic expression. However, the model's explainability analysis results (see Fig.~\ref{fig:non_nssi_weights}-\ref{fig:nssi_weights}) painted a more nuanced picture. In the non-NSSI state, the system exhibits a healthy capacity for introspection and self-correction: when precursors to somatization appear, a feedback pathway from \texttt{Somatization -> Unknown Factor 3 -> Defense Mechanisms} is activated, allowing the system to effectively "sense" and "process" this discomfort. In the NSSI state, however, this protective loop undergoes a \textbf{directional reversal}, becoming dominated by a maladaptive activation from \texttt{Defense Mechanisms -> Unknown Factor 3}. This "feedback loop reversal" finding perfectly corroborates our theory, proposed in the introduction, of NSSI as an "immature defense mechanism." It vividly demonstrates on a data level that, in the NSSI state, the defense mechanism itself is no longer an effective "problem-solver" but may instead be trapped in a state of "ineffective idling" or "creating noise," where its output fails to alleviate systemic pressure and instead exacerbates internal chaos.

This core finding also provides us with a new understanding of the potential of single-channel EEG analysis. The challenge of decoding complex mental states from single-channel EEG is often likened to \textbf{standing outside a massive stadium and trying to discern the cacophony of tens of thousands of spectators with just one microphone}. If a chaotic "football match" is underway inside, the task is nearly impossible. However, if a "concert" is taking place, with all spectators singing the same song in unison, even a single microphone can clearly capture the melody and rhythm. The methodology of this study, in concept, is precisely an attempt to facilitate this shift from a "football match" to a "concert." By presetting a "functional topology" model that aligns with clinical logic, we provide the GNN model with a "musical score" or a "grammatical framework," enabling it to identify \textit{pattern changes} against a structured background, rather than searching for details in noise. The "feedback loop reversal" captured by the model is that most critical "dissonant chord" or "key change"—the signal of the system's shift from order to chaos.

On a deeper level, this result also aligns with the Gestalt psychology and Level of Detail (LOD) theory we introduced in our theoretical construction \cite{tong2025perception}. The true state of the system is determined by the complete \textbf{"Gestalt"} formed by all nodes and pathways. The analysis performed by the model via GNNExplainer is precisely a computational LOD process: it does not get entangled in all the microscopic signal fluctuations but instead, through weight allocation, \textbf{"collapses" and identifies the macroscopic pattern change that is decisive for the overall configuration}—namely, the reversal of the key feedback loop. This provides strong computational evidence for our hypothesis that "a biased, simplified 'observational' method is needed to endow structure with meaning."

\subsection{Clinical Implications and Future Applications}
As a preliminary exploratory study, the clinical implications and future application potential of our findings may extend beyond the predictive performance of the model itself. We believe this work offers new perspectives and possibilities for the field of NSSI and the broader mental health domain on three levels.

\subsubsection{Mechanistic Understanding: A Shift from "Physical Localization" to "Functional Topology"}
First, this research provides a new theoretical framework for understanding the neural mechanisms of NSSI. Traditional neuroscience research has focused on physically localizing specific psychological functions to anatomical brain regions \cite{kosslyn2001neural}. While this paradigm has achieved great success, it still faces challenges in explaining complex, dynamic, and highly individualized mental phenomena like NSSI. We observe that function and structure do not always strictly correspond. Therefore, the core idea of this study is that, \textbf{compared to searching for a specific "brain region" where NSSI occurs, it may be more crucial to identify the specific "algorithm" or "syntax" that the brain's functional network follows during its occurrence}. Our "Functional-Energetic Topology Model" is precisely a theoretical abstraction and computational simulation of this "functional syntax." The research findings—particularly the revelation of the "key defense pathway failure" dynamic process—indicate that through this "de-physicalized" functional modeling, we may be able to capture dynamic transition patterns that are closer to the essence of psychological phenomena \cite{breakspear2017dynamic}.

\subsubsection{Objective Markers: Moving Towards Ecological Longitudinal Monitoring}
Second, this study demonstrates the immense potential of combining cutting-edge algorithms with consumer-grade portable EEG devices \cite{xue2024instrumentation}. The future of mental health assessment will inevitably require a \textbf{paradigm shift from the current reliance on "snapshot-like" subjective reports to an ecological, objective, and longitudinal monitoring conducted in the patient's real-life environment} \cite{powell2025exploring}. Just as wearable devices like the Apple Watch are revolutionizing the daily management of cardiovascular health, we believe that through efforts similar to this study, portable EEG devices will one day be able to provide continuous and meaningful biological markers for mental health.

The method we developed for analyzing EEG signals based on a GNN model is a significant step in this direction. It can not only assess the immediate effects of therapeutic interventions like meditation but also has the potential to be developed into an \textbf{objective biomarker for quantifying an individual's "Emotional Resilience."} For instance, by long-term monitoring of the health and stability of the key feedback regulatory loop discovered in this study (\texttt{Somatization -> Unknown Factor 3 -> Defense Mechanisms}), we might be able to quantify an individual's "self-correction capability" in the face of daily stress and identify periods of increased risk for decompensation.

\subsubsection{Early Warning Systems: The Core Engine of Future Digital Therapeutics (DTx)}
Finally, the ultimate vision of this research is to provide the core technological engine for the development of next-generation \textbf{Digital Therapeutics (DTx)} \cite{vasdev2024navigating}. Just as the emergence of large language models has shown us the immense power of large-scale data and advanced algorithms, we firmly believe that with sufficiently rich data collected from real-life settings, we can build more accurate and individualized decoders of mental states.

Our current model, although only a preliminary prototype, has already demonstrated the potential to be developed into a \textbf{real-time NSSI impulse warning system}. In the future, such a system could be integrated into a more comprehensive digital therapy application. When the model detects the "feedback loop reversal" decompensation pattern in the user's brain functional network in real-time, the system could proactively trigger an intervention, such as immediately guiding the user through a breathing exercise, playing a piece of relaxing music, or suggesting they contact a psychological counselor. This represents a paradigm shift from "reactive treatment" to "proactive prevention" and is what we believe to be the most exciting future that can be realized by combining the profound insights of a century of psychotherapy with 21st-century digital technology.

\subsection{Limitations of the Study}
Although this study offers a novel perspective and preliminary empirical evidence for understanding the neurodynamic mechanisms of NSSI, we must soberly acknowledge that, as an exploratory initial study, it has numerous inherent limitations. We candidly address these limitations here to define the applicable boundaries of our conclusions and to point the way for future research.

First, and most critically, is the \textbf{extremely small sample size}. This research is based on data from only three NSSI patients, which poses a severe challenge to the statistical power and generalizability of our conclusions. Although the leave-one-out cross-validation provides limited evidence for the model's generalization potential, we cannot rule out the possibility that the patterns discovered currently may only reflect the common characteristics of these three specific individuals. Future large-scale, multi-center studies are urgently needed, incorporating a more diverse range of NSSI patients with varying ages, genders, cultural backgrounds, and comorbidities, to test the universality of our findings.

Second, the use of a \textbf{single-channel, consumer-grade EEG device} constitutes another significant limitation of the study. The Fp1 single-point dry electrode device we employed, while enabling ecological daily data collection, does so at the cost of sacrificing spatial resolution and signal quality \cite{li2025tale}. We are unable to obtain information on the activity of other brain regions, which renders our constructed "functional topology network" entirely abstract at the neuroanatomical level. Furthermore, although we have made efforts to improve data quality and control for artifacts like muscle electricity through downsampling and algorithmic noise reduction, the inherent signal-to-noise ratio issues of consumer-grade devices may still pose a potential impact on the precision of the research findings.

Finally, we must profoundly reflect on the \textbf{theoretical abstraction of the model itself and its potential construction biases}. The seven-node topological structure proposed in this study largely originates from our clinical phenomenological observations and theoretical deductions, rather than being purely data-driven. For instance, the placement of the three "Other" nodes, while intended to simulate the complex transformation processes of information flow, still involves a degree of intuition and arbitrariness in their number and connection methods. On a deeper level, we must be wary of a potential "observer effect": the digital therapeutic paradigm we provided to the patients (such as rhythmic music meditation), along with the theoretical explanations used in our therapeutic communication, may have, to some extent, "shaped" or "guided" the patients' internal experiences and neural activity patterns to align with our theoretical framework \cite{pincus2024values}. The potential interaction between this research paradigm and the phenomenon being observed is a complex issue that needs to be cautiously addressed and deconstructed in future research.

In summary, we position this study as a \textbf{"Hypothesis-Generating"} work rather than a "Hypothesis-Confirming" one. As frontline clinicians, we hope that by sharing this still-immature but innovative exploration, we can provide researchers in the field with a starting point for thinking that differs from traditional paradigms and, by casting a brick to attract jade, inspire more rigorous and in-depth subsequent research.

\subsection{Future Research Directions}
Based on the preliminary findings and inherent limitations of this study, we have planned a series of future research directions aimed at continuously refining, validating, and expanding our proposed theoretical model and technical approach.

First, the most immediate and urgent task is to \textbf{overcome the current limitations in sample size and data quality}. We plan to conduct large-scale, multi-center studies to recruit a more diverse population of NSSI patients, in order to rigorously test the universality of the currently discovered "feedback loop reversal" pattern. Concurrently, we will introduce high-density, multi-channel EEG devices in future research. This will not only significantly improve the signal-to-noise ratio and spatial resolution but, more critically, it will enable us to attempt to establish preliminary correlations and mappings between our current abstract functional nodes and specific neural activity sources on the cerebral cortex (such as the frontal and parietal lobes), thereby building an exploratory bridge between the functional model and the brain's physical structure.

Second, we plan to expand our research perspective \textbf{from cross-sectional comparisons to longitudinal tracking and cross-disorder spectrum analysis}. By conducting long-term follow-ups with patients, we can observe how their functional topology networks dynamically evolve with therapeutic interventions, fluctuations in their condition, or aging. Furthermore, we believe that applying this study's analytical framework to patient populations with other psychiatric disorders (such as generalized anxiety disorder, major depressive disorder, and classic somatoform disorders) holds significant theoretical and clinical value. By comparing the similarities and differences in functional topology patterns across different diseases, we hope to answer a core question: is the dynamic pattern we have discovered specific to NSSI, or is it a shared core pathophysiological link among a broader range of emotion regulation disorders?

Finally, on the methodological front, we will also explore \textbf{more advanced and complex Graph Neural Network architectures}. Our current GCN model is a relatively basic framework; in the future, we could experiment with more advanced models like the Graph Attention Network (GAT), which can assign different weights to different neighboring nodes, to obtain more refined explainability. More futuristically, we envision evolving the current two-dimensional abstract topology into a \textbf{three-dimensional dynamic model embedded in a simulated physical space}. We speculate that by introducing interaction principles from physical systems, such as "gravity" and "repulsion," to simulate the interactions between nodes, we might be able to construct a dynamic system model that can better embody the holistic "Gestalt" nature of psychological phenomena. This will be our long-term goal in striving for a deeper integration of clinical phenomenological insights with the frontiers of computational science.

\section{Conclusion}
This research successfully proposed and preliminarily validated an innovative computational model based on functional topology theory to unveil the neurodynamics behind Non-Suicidal Self-Injury (NSSI). The findings indicate that the model can not only effectively distinguish NSSI states using single-channel EEG signals and possesses a degree of cross-subject generalization potential, but more importantly, its explainability analysis reveals that the occurrence of NSSI is critically associated with the dysfunction and directional reversal of a key feedback regulatory loop responsible for processing somatic sensations.

In conclusion, this exploratory work provides a novel, computable, and dynamic perspective for understanding the internal psychological processes of NSSI. It demonstrates the immense potential of integrating clinical theory with modern computational science to develop a new generation of objective assessment tools for mental health, thereby opening new avenues for both theoretical advancement and clinical practice in the field.

\section*{Funding}
This study was funded by the Science and Technology Plan Project of Inner Mongolia [Grant No. 2022YFSH0086] and the Inner Mongolia Medical University Joint Project [Grant No. YKD2023LH063]. We are grateful for their support.


\section*{Ethics Statement}
This study was ethically approved by the Ethics Review Committee of Inner Mongolia People’s Hospital. All par ticipants provided informed consent and their EEG data and evaluation records will be used for research while preserving anonymity. We adhere to ethical and research standards.

\section{Abbreviation}\label{secAl}

\begin{table}[h]
\centering
\caption{List of Abbreviations} 
\label{tab:abbreviations}
\footnotesize 
\setlength{\tabcolsep}{4pt}
\begin{tabular}{ll}
\toprule
\textbf{Abbr.} & \textbf{Full Name} \\ 
\midrule
BFRB  & Body-Focused Repetitive Behavior \\
dlPFC & Dorsolateral Prefrontal Cortex \\
DTx   & Digital Therapeutics \\
EEG   & Electroencephalography \\
EMG   & Electromyography (Frontal Muscle Activity) \\
EOG   & Electrooculography (Eye Movements) \\
ERP   & Event-Related Potential \\
FFT   & Fast Fourier Transform \\
GAT   & Graph Attention Network \\
GCN   & Graph Convolutional Network \\
GNN   & Graph Neural Network \\
ICA   & Independent Component Analysis \\
LOD   & Level of Detail \\
LOSOCV& Leave-One-Subject-Out Cross-Validation \\
LPP   & Late Positive Potential \\
NSSI  & Non-Suicidal Self-Injury \\
PFC   & Prefrontal Cortex \\
PSD   & Power Spectral Density \\
ReLU  & Rectified Linear Unit \\
TGAM  & ThinkGear ASIC Module \\
vmPFC & Ventromedial Prefrontal Cortex \\
XAI   & Explainable Artificial Intelligence \\
\bottomrule
\end{tabular}
\end{table}


\printbibliography 


\end{document}